\documentclass[%
 reprint,
 amsmath,amssymb,
 aps,
]{revtex4-2}

\usepackage{graphicx}
\usepackage{xcolor}
\usepackage{dcolumn}
\usepackage{bm}
\usepackage{physics}

\usepackage{enumitem}
\usepackage{float}
\usepackage{placeins}

\begin{document}

\title{Electron binding by charge disorder surrounding gapped bilayer graphene}

\author{Jude McClay$^{1,2}$, Angelika Knothe$^{3}$, Igor Rozhansky$^{1,2*}$, Vladimir Fal'ko$^{1,2,4}$}
\affiliation{$^1$National Graphene Institute, University of Manchester, Manchester M13 9PL, United Kingdom}
\affiliation{$^2$Department of Physics \& Astronomy, University of Manchester, Manchester M13 9PL, United Kingdom}
\affiliation{$^3$Institut f\"ur Theoretische Physik, Universit\"at Regensburg, D-93040 Regensburg, Germany}
\affiliation{$^4$Research Institute of Electrical Communication, Tohoku University, Sendai 980-8577, Japan}

\email{igor.rozhanskiy@manchester.ac.uk}

\date{\today}

\begin{abstract}

Bilayer graphene (BLG) is a unique electronic material offering electrostatic tunability of its band gap. Here, we study how charge disorder in the dielectric environment modifies band-edge states by binding electrons and holes to Coulomb centres, taking into account the in- and out-of-plane dielectric response of Bernal bilayers. We apply this framework to (a) hBN-encapsulated medium-size-gap BLG used for gate-defined quantum dots, and (b) BLG under double n/p ionic gating. 
In hBN/BLG/hBN, we find that single remote charges produce only weakly bound states, with binding energies generally small compared with the BLG gap.
Under ionic gating, we find that the bound states are too large to be trapped by individual ions -- localisation arises from fluctuations in the ion density, whose magnitude is limited by Coulomb charging energy at the freezing temperature $T_f$ of the ionic configuration. Thus, in both regimes, charge disorder does not produce deep midgap states, with the bound states remaining closer to the conduction and valence band edges.

\end{abstract}

\maketitle

Bernal-stacked bilayer graphene (BLG) features  electrically tuneable band gap~\cite{PhysRevLett.96.086805,Oostinga2008,Slizovskiy2021,Icking2022,PhysRevB.74.161403}, opened by an out-of-plane displacement field that controls an interlayer potential energy difference,  $\Delta$. This tunability underpins the use of BLG as a platform for electrostatically defined and controlled quantum dots and quantum wires~\cite{Allen2012,Overweg2018,PhysRevX.8.031023,PhysRevLett.123.026803,OverwegPRL2018} explored for spin/valley-qubits~\cite{Banszerus2023Nature,Hecker2023NatCommun,Garreis2024,Denisov2025}. While in such gate-controlled devices the feasible range of gaps extends up to $\sim100$ meV~\cite{Icking2022},  much larger gaps can be achieved by means of  ionic gating, where high density of ions is delivered electrolytically straight to the BLG surfaces~\cite{Efetov2011,melegari2026,Chakraborty2009,Chien2016,Grover2017,Mak2009,Ye2011}. 
\begin{figure}[!tb]
    \centering    \includegraphics[width=1\linewidth]{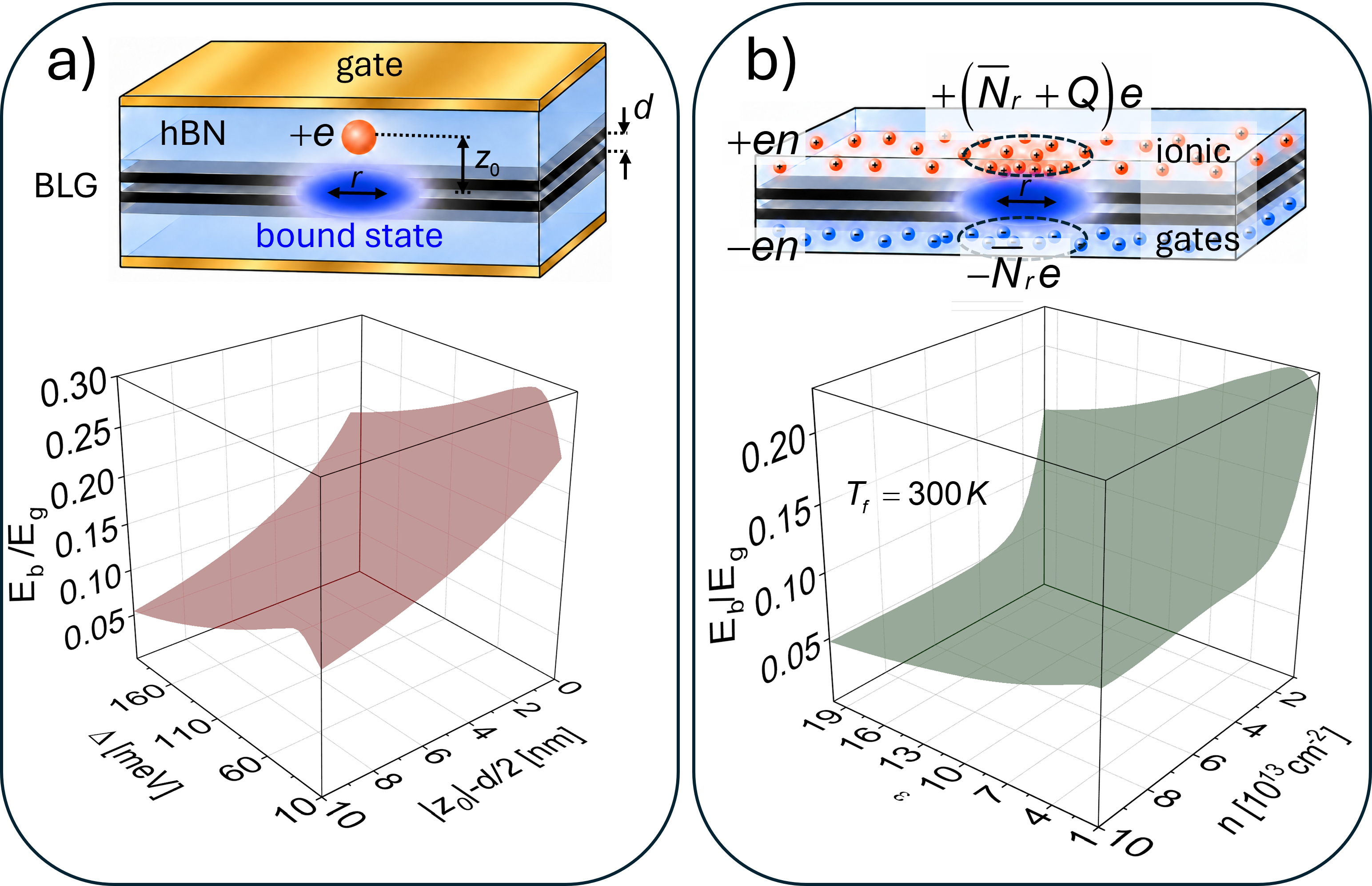}
    \caption{Two charge-disorder regimes in gapped BLG; $E_{\rm b}$ is the
binding energy and $E_g$ the spectral gap.
(a) hBN-encapsulated BLG, relevant to gate-defined quantum dots, with
$\varepsilon=4.5$ for hBN. The interlayer potential difference $\Delta$
is gate controlled; the lower panel shows $E_{\rm b}/E_g$ versus
$\Delta$ and the impurity distance from the nearest graphene layer.
(b) Double ionic gating. The mean ion density $n$ controls the
displacement field and the corresponding self-consistent $\Delta$.
For an effective bound-state radius $r$,
$\overline N_r=\pi n r^2$ is the mean number of ions on one interface
within the bound-state area, while $Q$ is the magnitude of the local fluctuation
in the difference between the ion numbers on the two interfaces.
The lower panel shows $E_{\rm b}/E_g$ at $T_f=300\,\mathrm{K}$,
where $T_f$ is the freezing temperature of the ionic configuration,
versus $n$ and the polymer-matrix dielectric constant $\varepsilon$.}
    \label{fig:schematic}
\end{figure}

As any other semiconductor, gapped BLG can host electronic states localised by  charged impurities embedded in the encapsulating environment (which can impede the use of gapped regions of BLG as insulating barriers~\cite{Martin2008,Zhang2009,Xue2011}), or ions implicit in electrolytic gating. As noted by several earlier studies~ \cite{NilssonCastroNeto2007,PhysRevB.89.041405,PhysRevB.96.165403,Gorbar2024,PhysRevB.98.115406}, tunabilities of the BLG band gap and 
band edge dispersion prescribe broad variations to spectral properties of such bound states. Those would also depend on the  distance from the BLG to the Coulomb centre, as well as on the size of the ionic charge. At the same time, intrinsic in-plane polarisability of bilayers is substantial, even when those are gapped~\cite{HwangDasSarma2008,WangChakraborty2010,TriolaRossi2012}, and 
it has been noticed~\cite{PhysRevB.96.165403,Gorbar2024} that this strongly reduces binding energy, shifting earlier suggested midgap states~\cite{PhysRevB.89.041405} towards band edges, also avoiding an 'atomic collapse' of donor states proposed in Ref.~\cite{PhysRevB.98.115406}.

Here, we extend the analysis of electron/hole binding by single or multiple ions in the  BLG environment with a view to describe two experimentally relevant systems sketched in Figure~\ref{fig:schematic}: (a) hBN-encapsulated BLG (such as used for gate-defined quantum circuits) perturbed by a rare single charged defect in  hBN placed at some distance $z_0$ from BLG, and (b) a BLG with large band gap set by double ionic gating, where the nearby ions themselves provide the Coulomb disorder. We analyse the resulting in-gap states (both ground states and excited states) using the full four-band Slonczewski--Weiss--McClure (SWMcC) Hamiltonian and including both the out-of-plane dielectric response of BLG~\cite{Slizovskiy2021} and its in-plane response within the random-phase approximation (RPA)~\cite{HwangDasSarma2008,WangChakraborty2010,TriolaRossi2012}. For both systems, (a) and (b), we find that the binding of electrons is relatively weak, with the binding energy of even the deepest state, $E_b$, being only a small fraction of the actual single-particle spectral gap of the BLG, as summarised in the bottom panels of Figure~\ref{fig:schematic}. 

{\bf For the gapped hBN-encapsulated BLG,} we consider a positive impurity of charge $+e$ at out-of-plane coordinate $z_0$, measured from the bilayer centre, in the  environment with the  dielectric constant $\varepsilon$, as shown in Figure~\ref{fig:schematic}a. The impurity produces an external Coulomb potential,
\begin{equation}
\label{eq:ionpot}
U_{\rm ext}^{t/b}(\mathbf r)=
-\frac{e^2}{4\pi\varepsilon_0\varepsilon}
\frac{\exp(-{\mathbf r}^2/L_g^2)}
{\sqrt{{\mathbf r}^2+(z_0\mp d/2)^2}},
\end{equation}
on the  top ($t$) and bottom ($b$) layers of the bilayer, separated by  $d=0.335\,\mathrm{nm}$, where ${\mathbf r}$ is the in-plane coordinate. A long-distance cutoff, at $L_g=20\,\mathrm{nm}$, is introduced to mimic screening by remote gates. Because the remote-impurity potential is smooth on the lattice scale, intervalley mixing is negligible; we therefore solve the bound-state problem in valley $K$, obtain the $K'$ spectrum by time reversal, and include both valleys in the screening calculation below.

The BLG Hamiltonian~\cite{PhysRevLett.96.086805} is written in the four-component sublattice basis $(A^{t},B^{t},A^{b},B^{b})$, as 
\begin{equation}
\label{eq:BLGham}
\begin{aligned}
    &\hat{H}^{0}_{K} =
    \begin{pmatrix}
        -\frac{\Delta}{2} + t_{g} g &  - \gamma_{0} f & \gamma_{4} f & - \gamma_{3} f^{*} \\
        -\gamma_{0} f^{*} & -\frac{\Delta}{2} + t_{g} g & \gamma_{1} & \gamma_{4} f \\
        \gamma_{4} f^{*} & \gamma_{1} & \frac{\Delta}{2} + t_{g} g & - \gamma_{0} f \\
        - \gamma_{3} f & \gamma_{4} f^{*} & - \gamma_{0} f^{*} & \frac{\Delta}{2} + t_{g} g
    \end{pmatrix}, \\[6pt]
    &f(\bm{k}) \equiv e^{i \frac{k_{y} a}{\sqrt{3}}} + 2 e^{-i \frac{k_{y} a}{2 \sqrt{3}}} \cos{\left(\frac{k_{x} a}{2} + \frac{2\pi}{3}\right)}, \\
    &g(\bm{k}) \equiv 3 + 2 \cos{\left( k_{x} a + \frac{4\pi}{3}\right)} \\
    & \quad \quad \quad \;\;\, +4 \cos{\left( \frac{k_{y} \sqrt{3} a}{2} \right)} \cos{\left( \frac{k_{x} a}{2} + \frac{2\pi}{3} \right)},
\end{aligned}
\end{equation}
where $a=0.246\,\mathrm{nm}$ is the graphene's lattice constant and $\hbar\bm{k}$ is the valley momentum. This SWMcC Hamiltonian is parametrised~\cite{McCannKoshino2013} using  $\gamma_{0}=3.15\,\mathrm{eV}$, $\gamma_{1}=0.38\,\mathrm{eV}$, $\gamma_{3}=\gamma_{1}$, $\gamma_{4}=0.14\,\mathrm{eV}$ and $t_g=0.1\,\mathrm{eV}$, where $t_g$ is the intralayer next-nearest-neighbour hopping included to capture a weak electron--hole asymmetry. The interlayer asymmetry is represented by layer-dependent energies $\mp\Delta/2$.

\begin{figure*}[!t]
    \centering
\includegraphics[width=0.96\linewidth]{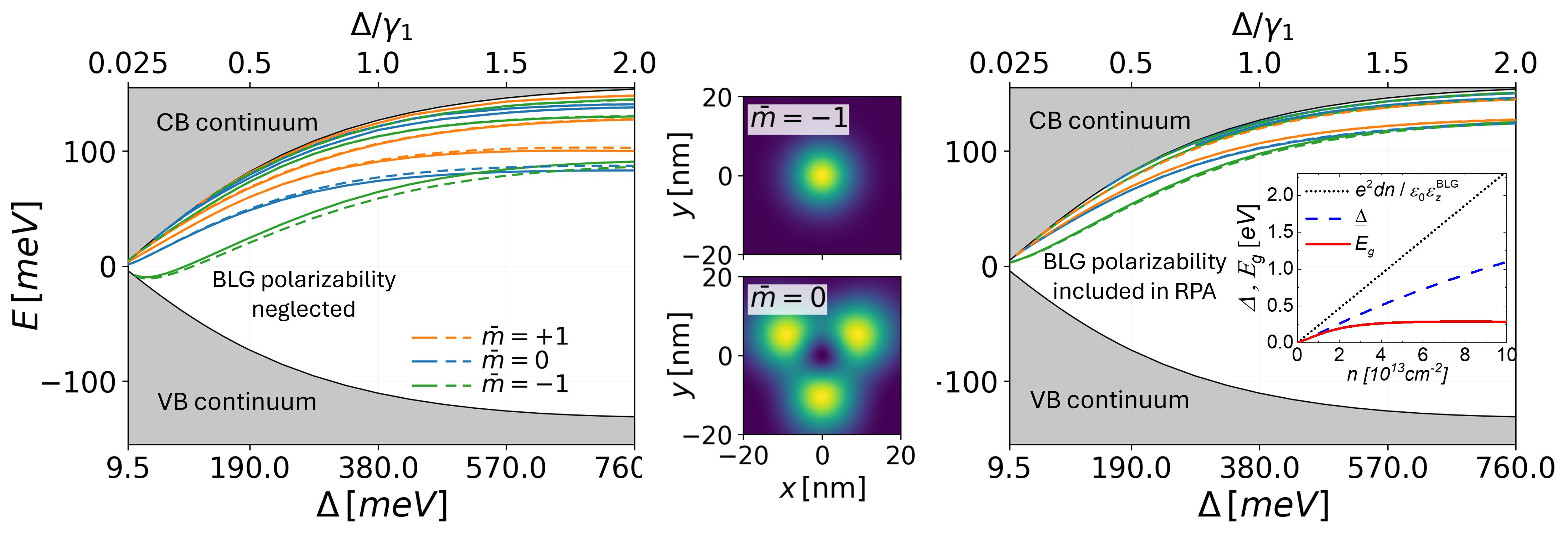}
    \caption{Screening shifts bound levels towards the band edges. The valley-$K$ spectrum is shown for an impurity $1\,\mathrm{nm}$ from the nearest graphene layer. The left plot uses the external potential only; the middle plots show representative bound-state probability densities with different spatial symmetries at $\Delta=100\,\mathrm{meV}$; the right plot includes out-of-plane and in-plane RPA screening. Solid and dashed curves correspond to impurities above and below the bilayer, respectively; the continua are shaded grey. The inset compares the ionic displacement-field contribution (first term of Eq.~\ref{eq:Delta_ionic}), the self-consistent $\Delta$, and $E_g$ for double ionic gating.}
    \label{fig:spectrum}
\end{figure*}
 We describe the bound states by slowly varying envelope functions modulating the microscopic Bloch states represented~\cite{PhysRevB.101.235423,KnotheFalco2018} using a  harmonic-oscillator basis, $\psi_{\eta,m}(\bm r)$, 
\begin{align*}
&\ket{\Psi_{K}} 
= \sum_{X}  \sum_{\eta,m}  c^{X}_{\eta,m}\,\psi_{\eta,m}\ket{X_{K}}, \\
&\psi_{\eta,m} 
= \frac{e^{im\theta}}{\sqrt{2\pi \mathcal{N}_{\eta,m}}}
\left(\frac{{\mathbf r^2}}{\ell_0^2}\right)^{\frac{|m|}{2}}
e^{-{\mathbf r^2}/2\ell_0^{2}}
L^{|m|}_{(\eta-|m|)/2}
\!\left(\frac{{\mathbf r^2}}{\ell_0^{2}}\right).
\end{align*}
Here, $\ket{X_{K}}$ denotes the Bloch-function component on sublattice $X$ evaluated at the centre of valley $K$; $\eta=0,1,2,\ldots$; $m=-\eta,-\eta+2,\ldots,\eta$, ; $\mathcal{N}_{\eta,m}=\frac{\ell_0^{2}}{2}\frac{\left((\eta+|m|)/2\right)!}{\left((\eta-|m|)/2\right)!}$; $L_n^{|m|}$ are associated Laguerre polynomials. The angular momentum $m$ is referenced to the dimer sites $A^{b}$ and $B^{t}$ and is shifted by $\pm1$ for the non-dimer components. As a result, the Hamiltonian couples $\ket{A^{t},m-1}\leftrightarrow\ket{B^{t},m}\leftrightarrow\ket{A^{b},m}\leftrightarrow\ket{B^{b},m+1}$. Note that $\hat H^0_K$ in Eq.~\ref{eq:BLGham} reflects the $C_3$ symmetry of the lattice leading to trigonally warped bands and dictating the mixing of $m$ indices so that only the reduced angular-momentum quantum number $\bar m=m\;\mathrm{mod}\;3$ is conserved, where $\bar m=0,\pm1$. We therefore organise the basis into these three sectors with the Hamiltonian for each sector diagonalised independently. In the performed numerical simulation, the basis is truncated at a maximum value of $\eta$, which is increased until the energy spectrum in each sector is numerically converged.
We characterise spatial extents of bound states by the effective radius
\begin{equation}
\label{eq:IPR}
r=[\pi\int d^2\mathbf r'\,p^2(\mathbf r')]^{-1/2};\quad p(\mathbf r)=\sum_X|\Psi_X(\mathbf r)|^2,
\end{equation}
where the sum runs over the four sublattices, and $\Psi_X(\mathbf r)=\sum_{\eta,m}c^X_{\eta,m}\psi_{\eta,m}(\mathbf r)$.
We also adapt the oscillator length $\ell_0$ to the characteristic size of the bound state to accelerate  convergence.

In the implemented  numerical procedure the sets of coefficients $c^{X}_{\eta,m}$ are found  as eigenvectors of the Hamiltonian matrix,
\begin{widetext}
\begin{equation}
\label{eq:fullham}
\begin{aligned}
    \bra{X'_{K},\eta',m'} \hat H \ket{X_{K}, \eta, m}  = 
    & \int \frac{d^{2} \bm{k}}{(2 \pi)^{2}} \left( \bra{X'_{K}} \hat{H}^{0}_{K} \ket{X_{K}} \int d^{2}\bm{r} d^{2}\bm{r}' \psi_{\eta',m'}^{*}(\bm{r}') e^{i \bm{k} \cdot (\bm{r}'-\bm{r})} \psi_{\eta,m}(\bm{r}) \right) \\
    & + \int d^{2}\bm{r} \, \bra{X'_{K}}
    \begin{pmatrix}
        U^{t}(\mathbf r) & 0 & 0 & 0 \\
        0 & U^{t}(\mathbf r) & 0 & 0 \\
        0 & 0 & U^{b}(\mathbf r) & 0 \\
        0 & 0 & 0 & U^{b}(\mathbf r) \\
    \end{pmatrix} \ket{X_{K}} \, \psi_{\eta',m'}^{*}(\bm{r}) \psi_{\eta,m}(\bm{r}).
\end{aligned}
\end{equation}
\end{widetext}
Here, $U^{t/b}$ denotes either the external potential of Eq.~\ref{eq:ionpot} or the corresponding screened potential, see below for details. To speed up the calculation we evaluate the $k$-space integral by Gauss--Laguerre quadrature. 
\begin{figure}[!tbp]
    \centering
    \includegraphics[width=0.97\linewidth]{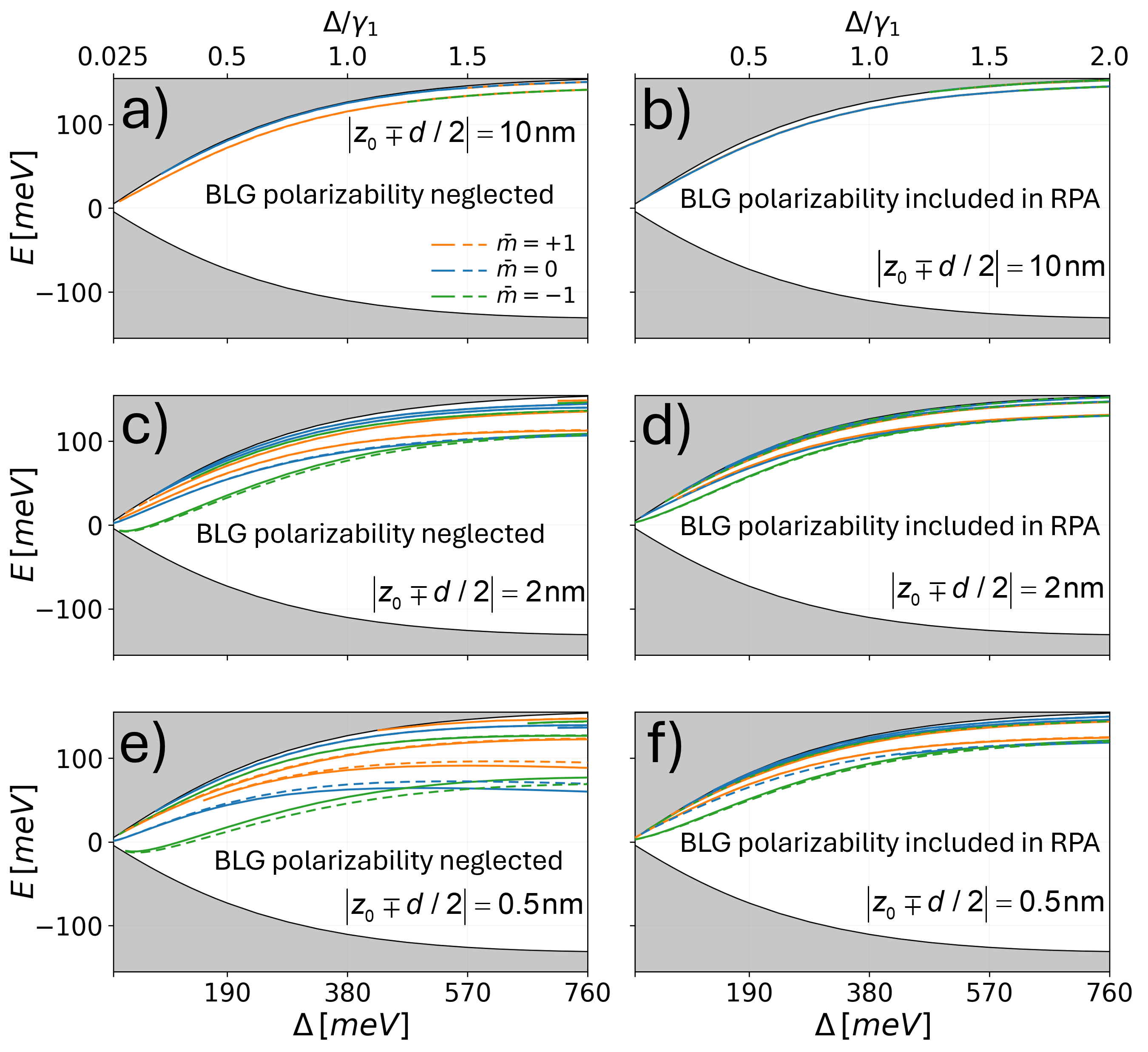}
    \caption{Binding weakens with impurity--bilayer separation and screening. Valley-$K$ spectra are shown for impurity distances of $10\,\mathrm{nm}$ (a,b), $2\,\mathrm{nm}$ (c,d), and $0.5\,\mathrm{nm}$ (e,f) from the nearest layer. Left: external potential only; right: out-of-plane and in-plane RPA screening. Solid and dashed lines correspond to impurities above and below the bilayer; colours distinguish $\bar m=+1,0,-1$.}
    \label{fig:spectrumdistances}
\end{figure}

We further account for screening by BLG, which has been shown to be important
~\cite{HwangDasSarma2008,WangChakraborty2010,TriolaRossi2012}.
First, to account for the intrinsic out-of-plane dielectric response of BLG, we follow
Ref.~\cite{Slizovskiy2021} and use $\varepsilon_z^{\rm BLG}=2.6$ determined by the out of plane polarisability of monolayer graphene.
This leaves the layer-symmetric part of the external potential unchanged and
reduces the layer-antisymmetric part by
$1/\varepsilon_z^{\rm BLG}$.
The resulting potentials on the two graphene layers are
\begin{equation}\label{eq:outofplane}
\widetilde U_{t(b)}
=
\frac{1}{2}\left[
U_t^{\rm ext}+U_b^{\rm ext}
\pm
\frac{U_t^{\rm ext}-U_b^{\rm ext}}
{\varepsilon_z^{\rm BLG}}
\right],
\end{equation}
where the upper (lower) sign corresponds to the top (bottom) layer.

Second, we include in-plane electronic screening by the BLG bands within the
layer-resolved RPA~\cite{HwangDasSarma2008,WangChakraborty2010}.
For this we need to consider 
the Forier transforms
$\widetilde{\mathbf U}(\mathbf q)
=(\widetilde U_t(\mathbf q),\widetilde U_b(\mathbf q))^T$
of the layer potentials obtained after the out-of-plane
correction in Eq.~\ref{eq:outofplane}. The fully screened potential in
momentum space is then
\begin{align}
&\mathbf U^{\rm scr}(\mathbf q)
=
\left[\hat I+\hat V(\mathbf q)\hat\Pi(\mathbf q)\right]^{-1}
\widetilde{\mathbf U}(\mathbf q),
\,\,\,
\hat\Pi(\mathbf q)=
\begin{pmatrix}
\Pi_{tt} & \Pi_{tb}\\
\Pi_{bt} & \Pi_{bb}
\end{pmatrix},
\nonumber\\
&\Pi_{\lambda\lambda'}(\mathbf q)
=
\sum_{\xi,\nu,\nu',\mathbf k}
\frac{
n_F(E^\xi_{\nu,\mathbf k})
-n_F(E^\xi_{\nu',\mathbf k+\mathbf q})
}{
E^\xi_{\nu',\mathbf k+\mathbf q}
-E^\xi_{\nu,\mathbf k}
}
\mathcal F^{\lambda\lambda'}_{\xi\nu\nu'}
(\mathbf k,\mathbf q),
\nonumber\\
&\mathcal F^{\lambda\lambda'}_{\xi\nu\nu'}(\mathbf k,\mathbf q)
=
\langle u^\xi_{\nu,\mathbf k}|
\hat P_\lambda
|u^\xi_{\nu',\mathbf k+\mathbf q}\rangle
\langle u^\xi_{\nu',\mathbf k+\mathbf q}|
\hat P_{\lambda'}
|u^\xi_{\nu,\mathbf k}\rangle, 
\label{eq:screening}
\end{align}
where 
\begin{equation*}\label{eq:layer_coulomb}
\hat V(\mathbf q)
=
\frac{e^2}{2\varepsilon_0\varepsilon q}
\left[
\begin{aligned}
&
\begin{pmatrix}
1 & e^{-qd}\\
e^{-qd} & 1
\end{pmatrix}\\
&-
\frac{\left(\varepsilon_z^{\rm BLG}-1\right)(1-e^{-qd})}
{4\varepsilon_z^{\rm BLG}}
\begin{pmatrix}
1 & -1\\
-1 & 1
\end{pmatrix}
\end{aligned}
\right];
\end{equation*}
 $n_F$ is the Fermi function; $\hat I$ is the $2\times2$ identity matrix,
and $\hat P_t=\mathrm{diag}(1,1,0,0)$ and
$\hat P_b=\mathrm{diag}(0,0,1,1)$ are the wavefunction projectors onto sublattices in the top and bottom
layers, respectively. The indices $\lambda,\lambda'=(t,b)$ label the layers,
$\xi=K,K'$ labels the two valleys, and $\nu,\nu'$ run over the four BLG bands; 
$\sum_{\mathbf k}\equiv2\int d^2k/(2\pi)^2$;
 $E^\xi_{\nu,\mathbf k}$ and
$|u^\xi_{\nu,\mathbf k}\rangle$ are the energies and four-component
eigenstates of the unperturbed BLG Hamiltonian
$\hat H^0_\xi$ in valley $\xi$, respectively, as defined in Eq.~\ref{eq:BLGham}.

At charge neutrality, Fermi level lies in the gap, so the static
response is dominated by interband processes rather than free-carrier
intraband screening. Trigonal warping gives
$\Pi_{\lambda\lambda'}(\mathbf q)$ a $C_3$ angular dependence, which we
retain. In what follows, ``screened'' denotes calculations including both the
out-of-plane dielectric response and the in-plane RPA screening. The resulting
screened layer potentials $U^{\rm scr}_{t/b}(\mathbf r)$, obtained by Fourier
transforming $\mathbf U^{\rm scr}(\mathbf q)$ back to real space, are then used
in Eq.~\ref{eq:fullham}. Further details of the layer-resolved polarisation,
matrix RPA screening, and numerical convergence are given in Sec.~1 of the
Supporting Information, in a qualitative agreement with earlier publications~\cite{HwangDasSarma2008,WangChakraborty2010,TriolaRossi2012}.

\begin{figure}[tbp]
    \centering
\includegraphics[width=0.98\linewidth,height=0.36\textheight,keepaspectratio]{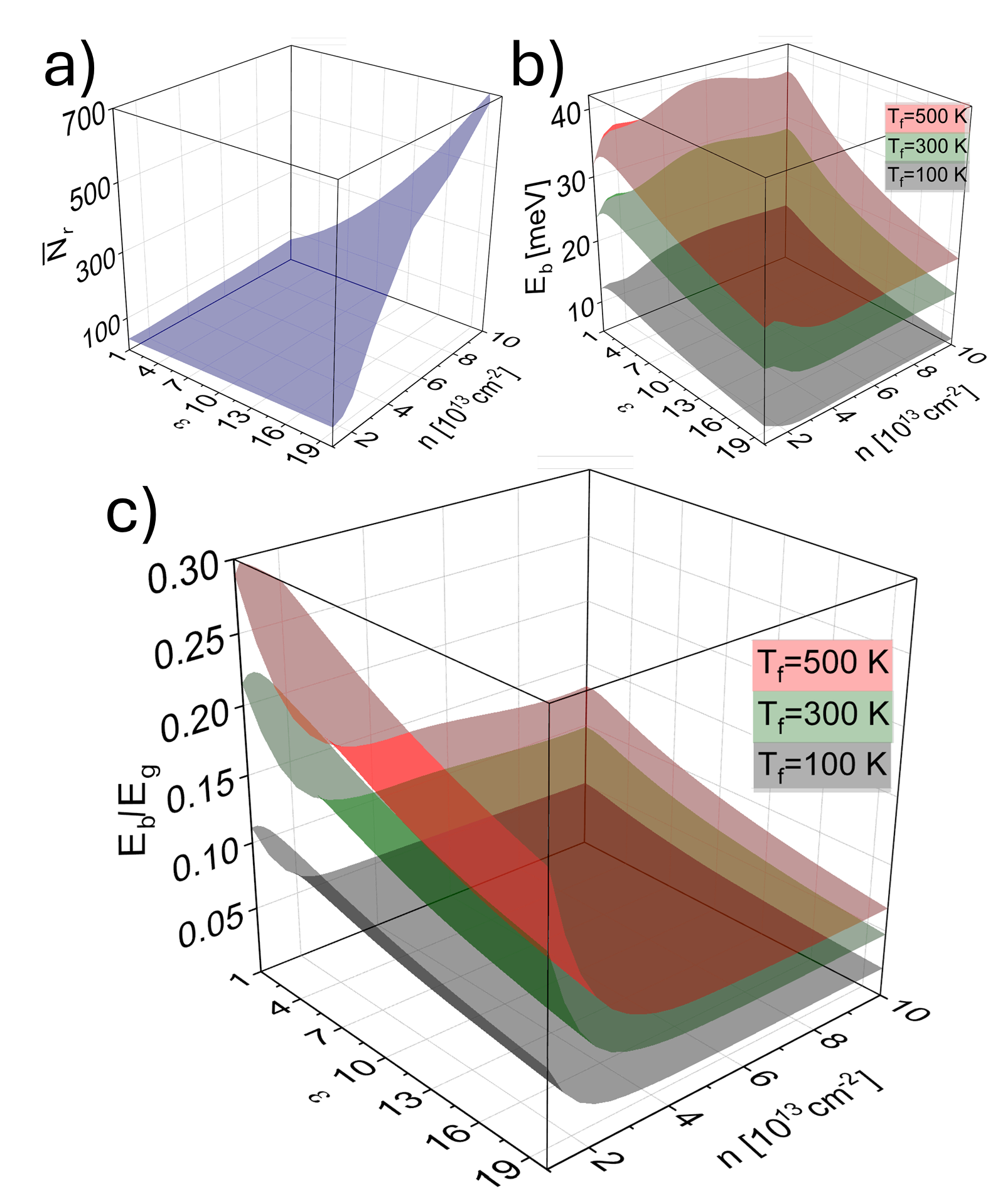}
    \caption{The temperature $T_f$ at which the ionic configuration freezes sets the characteristic ionic-charge fluctuation $Q_T$ through the charging-energy condition $E_{\rm ch}(Q_T)=k_{\rm B}T_f$. For ions $0.5\,\mathrm{nm}$ from the nearest graphene layer, (a) the mean ion count $\overline{N}_r$ at $Q=1$, (b) $E_{\rm b}$ at $Q_T$, and (c) $E_{\rm b}/E_g$ are shown versus $n$ and the polymer-matrix dielectric constant $\varepsilon$. Panels (b,c) compare $T_f=100$, $300$, and $500\,\mathrm{K}$.}
    \label{fig:ionicmap}
\end{figure}

For hBN-encapsulated BLG, the external gates control the displacement field
and hence the interlayer asymmetry $\Delta$, which we use to parameterise
the calculations. 
Figure~\ref{fig:spectrum} shows the bound-state spectrum in valley $K$ for $\bar m=0,\pm1$ versus $\Delta$, for an impurity $1\,\mathrm{nm}$ from the nearest graphene layer. Each state is spin and valley degenerate, but with inverted angular-momentum labels $\bar m\to-\bar m$ in valley $K'$ as compared to valley $K$. The computed spectra span the range relevant to both conventional and ionic gating. The middle panels in Figure~\ref{fig:spectrum} illustrate the distinct real-space probability-density patterns of representative bound states. Without screening, as $\Delta$ is reduced, the lowest $\bar m=-1$ state approaches the valence-band edge and eventually merges into the continuum. Screening preserves this evolution but shifts the bound levels substantially towards the band edge. Figure~\ref{fig:spectrumdistances} shows the same trend with impurity distance: increasing the separation weakens the binding. At high ionic density, the self-consistent $\Delta$ and the spectral gap $E_g$ differ markedly, as shown in the inset of Figure~\ref{fig:spectrum}. Since $E_g$ is the actual conduction--valence band separation, we use $E_{\rm b}/E_g$ to quantify the bound-state depth in Figures~\ref{fig:schematic} and \ref{fig:ionicmap}c.

\begin{figure}[tbp]
    \centering    \includegraphics[width=0.97\linewidth]{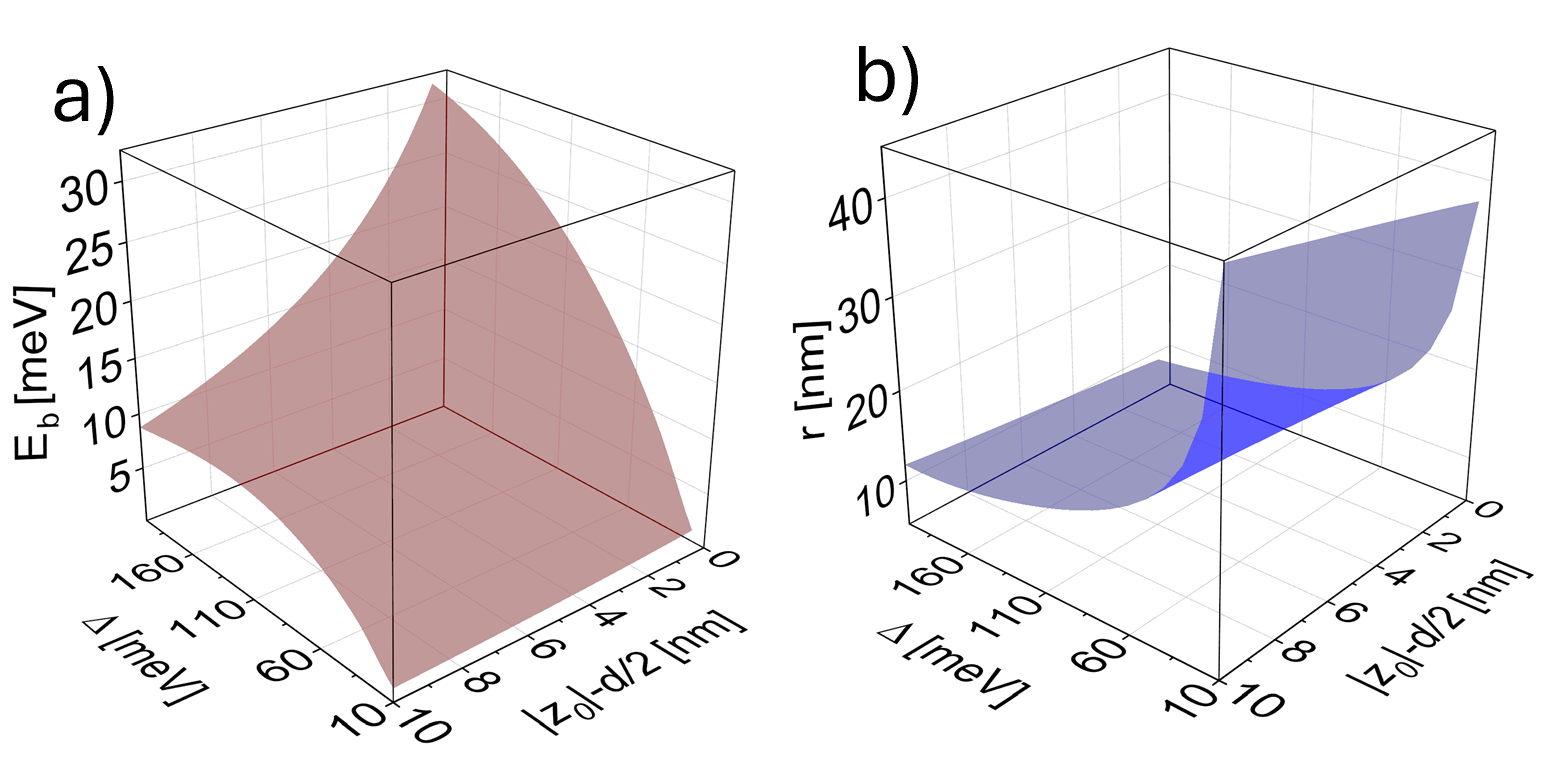}
    \caption{With screening, bound states remain shallow over most of the parameter range. (a) Binding energy $E_{\rm b}$ and (b) radius $r$ versus $\Delta$ and impurity distance $|z_0|-d/2$ from the nearest graphene layer.}
    \label{fig:cube}
\end{figure}

Figure~\ref{fig:cube} shows $E_{b}$ and the effective bound-state radius $r$ over the $(\Delta,|z_0|-d/2)$ range. With screening, the bound states produced by a single remote impurity remain shallow throughout the parameter range, which is highlighted in Figure~\ref{fig:schematic}a, where we normalise the binding energy, $E_b$, by the actual spectral gap, $E_g$. This provides a parameter-dependent disorder scale for gate-defined BLG quantum dots~\cite{Allen2012,PhysRevX.8.031023,PhysRevLett.123.026803,Banszerus2023Nature,Hecker2023NatCommun,Garreis2024,Denisov2025}: size-quantised dot levels placed deeper in the gap than the corresponding impurity-bound state should suppress resonant leakage through such states, which is particularly relevant to BLG-based spin/valley qubits.

{\bf For BLG under double ionic gating,} as shown in Figure~\ref{fig:schematic}b, the spectral gap is controlled
by the charge densities, $+en$ and $-en$, of ions above and below BLG, respectively, which set the interlayer potential offset $\Delta$. To describe this potential offset, we use a self-consistent model~\cite{Slizovskiy2021} that accounts for out-of-plane polarisability of graphene planes,
\begin{equation}
\label{eq:Delta_ionic}
\Delta=
\frac{e^2dn}{\varepsilon_0\varepsilon_z^{\rm BLG}}
-\frac{e^2d}{4\varepsilon_0}
\left(1+\frac{1}{\varepsilon_z^{\rm BLG}}\right)(n_b-n_t).
\end{equation}
Here, $n_b-n_t$ is the electronic layer-density imbalance computed with the Hamiltonian~(\ref{eq:BLGham}), that incorporates the same value of $\Delta$.  The spectral gap $E_g$
is obtained from the separation between the conduction- and valence-band
edges of the resulting four-band spectrum, and it differs from the value of $\Delta$ being limited by the interlayer hopping $\gamma_1$.
To reach the BLG band gap $E_g$ in excess of 150 meV, one would need to use ion densities $n>10^{13}$ cm$^{-2}$ as shown in the inset in the right-hand panel of Figure~\ref{fig:spectrum}. 

Having analysed bound-state wavefunctions for those and higher densities and gaps
(see Sec.~2 of the Supporting Information), we noted that those always cover a large number of ions,  $\overline{N}_r=\pi n r^2 \gg 1$; see Figure~\ref{fig:ionicmap}a, with $r$ defined in Eq.~\ref{eq:IPR}. Therefore, binding of electrons can be provided only by fluctuations of the ionic density. To describe such a binding, we extend the analysis onto multiple-charged Coulomb centers, $Qe$, by substituting the confinement potential in Eq.~\ref{eq:ionpot} with  $U_{\rm ext}^{(Q)}=Q U_{\rm ext}^{(1)}$. 
Next, we consider $Q\sim \sqrt{2\overline{N}_r}$ as a number of excess ions above BLG as compared to underneath, covered by the wavefunction with a size $r(Q)$ of a bound state with the energy $E_b(Q)$ computed for that value of $Q$. 

Moreover, one needs to take into account an energy cost of a dense fluctuation of ions.
With a view that ion configuration is frozen at some freezing temperature $T_f$, we should use a constraint on the size of $Q$ such that  
\begin{equation}
\label{eq:ionic_thermal}
\frac{e^2Q^2}{4\pi\varepsilon_0\varepsilon r(Q)}\lesssim k_{\rm B}T_f.
\end{equation}
The latter constraint defines the limiting condition for the binding energy $E_b(Q)$ computed for electrons (or holes) for the maximum value of $Q$ for each given input ion density and dielectric constant of ion-hosting polymer matrix. 

In Figure~\ref{fig:ionicmap} we illustrate these limiting values for three
freezing temperatures of the ionic configuration,
$T_f=100$, $300$, and $500\,\mathrm{K}$. In particular, for ionic gating prepared at room temperature and quickly frozen for transport measurements, the maximum value of $T_f$ would be 300 $K$, which we use in the plot in Figure~\ref{fig:schematic}b, which shows that  
the bound states remain shallow across the entire parameter range:  the largest value $E_b/E_g\approx0.22$ occurs only where $\varepsilon\sim 1$, which is unlikely to be realised in real polymer environments~\cite{melegari2026}.

In summary, for hBN-encapsulated BLG, we find that a single remote charge produces only shallow bound states over the relevant parameter range. Their depth is controlled by the local gap and impurity distance, as mapped in Figure~\ref{fig:cube}: placing size-quantised dot levels sufficiently deep in the gap should therefore avoid resonant coupling to impurity-induced states. For ionically gated BLG, the bound states are too extended for localisation by a single ion. Instead, localisation arises from fluctuations of the ionic charge within the bound-state area, whose magnitude is limited by Coulomb charging energy when the ionic configuration freezes. Our calculations show that, over the experimentally relevant parameter range, these states can only be weakly bound (no mid-gap states), which justifies the descriptions of electronic transport in ion-gated BLG in terms of scattering of the carriers from weak charge disorder~\cite{Koshino_2009,PhysRevB.82.081407,PhysRevB.84.041406,Abergel2012}.

\section*{Acknowledgments}
A.K. acknowledges support from the Deutsche Forschungsgemeinschaft (DFG, German Research Foundation) within DFG Individual Grants KN 1383/4, KN 1383/7, and SFB 1277 Project A07 (Project-ID 314695032); J.McC. from CDT Graphene-NOWNANO, and V.F. from the British Council - ISPF grant 1185409051. We also thank Tohoku University Physics Department for hospitality. 

\bibliography{references}

\end{document}